\documentclass[twocolumn]{aastex631}

\shorttitle{Mono-enriched Stars}
\shortauthors{Hirai et al.}
\begin{document}
\title{SIRIUS: Identifying Metal-poor Stars Enriched by a Single Supernova in a Dwarf Galaxy Cosmological Zoom-in Simulation Resolving Individual Massive Stars}
\correspondingauthor{Yutaka Hirai}
\email{yutaka.hirai@astr.tohoku.ac.jp}

\author[0000-0002-5661-033X]{Yutaka Hirai}
\altaffiliation{JSPS Research Fellow}
\affiliation{Astronomical Institute, Tohoku University,
6-3 Aoba, Aramaki, Aoba-ku, Sendai, Miyagi 980-8578, Japan}

\author[0000-0001-8226-4592]{Takayuki R. Saitoh}
\affiliation{Department of Planetology, Graduate School of Science, Kobe University, 1-1 Rokkodai-cho, Nada-ku, Kobe, Hyogo 657-8501, Japan}

\author[0000-0002-6465-2978]{Michiko S. Fujii}
\affiliation{Department of Astronomy, Graduate School of Science, The University of Tokyo, 7-3-1 Hongo, Bunkyo-ku, Tokyo 113-0033, Japan}

\author{Katsuhiro Kaneko}
\affiliation{Department of Astronomy, Graduate School of Science, The University of Tokyo, 7-3-1 Hongo, Bunkyo-ku, Tokyo 113-0033, Japan}

\author[0000-0003-4573-6233]{Timothy C. Beers}
\affiliation{Department of Physics and Astronomy, University of Notre Dame,
225 Nieuwland Science Hall, Notre Dame, IN 46556, USA}
\affiliation{Joint Institute for Nuclear Astrophysics, Center for the Evolution of the Elements (JINA-CEE), East Lansing, MI 48824, USA}

\begin{abstract}
Metal-poor stars enriched by a single supernova (mono-enriched stars) are direct proof (and provide valuable probes) of supernova nucleosynthesis. Photometric and spectroscopic observations have shown that metal-poor stars have a wide variety of chemical compositions;  the star's chemical composition reflects the nucleosynthesis process(es) that occurred before the star's formation. While the identification of mono-enriched stars enables us to study the ejecta properties of a single supernova, the fraction of mono-enriched stars among metal-poor stars remains unknown. {Here we identify mono-enriched stars in a dwarf galaxy cosmological zoom-in simulation resolving individual massive stars.} We find that the fraction of mono-enriched stars is higher for lower metallicity, stars with [Fe/H] $< -2.5$. The percentages of mono-enriched stars are 11\% at [Fe/H] = $-$5.0 and 1\% at [Fe/H] = $-$2.5, suggesting that most metal-poor stars are affected by multiple supernovae. We also find that mono-enriched stars tend to be located near the center of the simulated dwarf. Such regions will be explored in detail in upcoming surveys such as the Prime Focus Spectrograph (PFS) on the Subaru telescope.
\end{abstract}

\section{Introduction} \label{sec:intro}
Chemical abundances of metal-poor (MP) stars reflect the nucleosynthesis process(es) that took place prior to their formation. The nucleosynthetic yields of supernovae (SNe) depend on a number of variables, including stellar mass, stellar evolution, rotation, and explosion mechanisms \citep[e.g.,][]{Heger2010, Nomoto2013, Tsiatsiou2024}. We can observationally constrain SNe nucleosynthesis using the chemical abundances of MP stars enriched by a single SN (mono-enriched stars). Thus, the identification of mono-enriched stars is critical for acquiring this information.

In the past few decades, {spectroscopic and photometric} observations have measured the chemical abundances of MP stars \citep[e.g.,][]{Beers1985, Beers1992, Beers2005, Tolstoy2009, Frebel2015}. \citet{Beers2005} classified MP stars according to their [Fe/H]\footnote{[Fe/H]=$\log(N_{\rm{Fe}}/N_{\rm{H}})-\log(N_{\rm{Fe}}/N_{\rm{H}})_{\sun}$, where $N_{\rm{Fe}}$ and $N_{\rm{H}}$ are the number densities of Fe and H, respectively.}: MP for [Fe/H] $<-1$, very metal-poor (VMP) for [Fe/H] $<-2$, extremely metal-poor (EMP) for [Fe/H] $<-3$, ultra metal-poor (UMP) for [Fe/H] $<-4$, and hyper metal-poor (HMP) for [Fe/H] $<-5$. The HK objective-prism survey reported spectroscopy of 1044 stars, 446 of which were classified as VMP \citep{Beers1985, Beers1992};
The Hamburg-ESO Survey (HES; \citealt{Christlieb2008}) carried out similar objective-prism observations, and supplied targets for high-resolution studies of additional VMP stars.  High-resolution spectroscopic follow-up of stars from the HK survey and the HES  have been reported by a number of previous authors (see, e.g., \citealt{Roederer2014}, and references therein).  Other high-resolution follow-up efforts continue to expand the list of VMP stars with detailed abundance information, based on stars identified from massive spectroscopic surveys, such as the SDSS \citep{York2000}, RAVE \citep{Steinmetz2006}, SDSS/SEGUE survey \citep{Yanny2009, Rockosi2022}, the LAMOST survey \citep{Deng2012}, the GALAH survey \citep{DeSilva2015}, 
SDSS/APOGEE \citep{Majewski2017}, the Gaia-ESO survey \citep{Gilmore2022}, and DESI \citep{Cooper2023}.  

Recent massive photometric surveys, in particular those that include narrow-band filters, such as Pristine \citep{Starkenburg2017}, J-PLUS \citep{Cenarro2019}, 
S-PLUS \citep{MendesdeOliveira2019}, SkyMapper \citep{Onken2019}, and SAGES \citep{Fan2023}, hold the promise to enormously expand the list of VMP targets for future follow-up at high spectral resolution.
It is important to identify mono-enriched stars based on these observations.  

The chemical abundances of some MP stars indicate that the stars are mono-enriched \citep[e.g.,][]{Ishigaki2018, Aoki2014, Jeong2023, Ji2024}. \citet{Aoki2014} found that the VMP star SDSS J001820.5-093939.2 has a low [$\alpha$/Fe] ratio and an odd-even abundance pattern, suggesting that it was formed from the ejecta of SNe of stars more massive than 140 M$_{\sun}$. \citet{Ji2024} reported the lowest known ratios of [N/Fe], [Na/Fe], [K/Fe], [Sc/Fe], [Ba/Fe], and a clear contrast between odd and even elements for this star. Although none of the models can fully explain the chemical abundance pattern, they prefer the existence of a SN progenitor of $>$50 $M_{\sun}$. 

Carbon-enhanced metal-poor (CEMP) stars could also be mono-enriched (e.g., \citealt{Beers2007, Lee2013, Yoon2016, Yoon2019}, and references therein). The small abundance dispersions (particularly for C) reported in chemo-dynamically tagged groups (CDTGs) by \citet{Zepeda2023}  for CEMP stars classified as Group II stars in the Yoon-Beers diagram \citep{Yoon2016} could be signatures of enrichment by rotating massive stars \citep[e.g.,][]{Meynet2006} or faint SNe \citep[e.g.,][]{Ishigaki2014}.

Estimates of the fractions of mono-enriched stars as a function of the metallicity could help us identify such stars. \citet{Hartwig2023} applied machine learning to estimate the number of SNe contributed to the chemical abundances of previously observed EMP stars. They found that 31.8\% of 462 EMP stars are mono-enriched, and that the mono-enriched fraction decreases toward higher metallicity. However, the machine-learning approach relies on the predicted nucleosynthetic yields as well as the observed data, both of which can carry large uncertainties. Numerical simulations with detailed treatment of metal dilution provide a complimentary  approach, and can estimate the mono-enriched fraction more directly.

Galaxy simulations that can resolve individual stars (star-by-star simulations) could be used for this purpose \citep[e.g.,][]{Emerick2018, Hu2019, Lahen2020, Gutcke2021, Hirai2021, Calura2022, Calura2024,  Brauer2024, Deng2024, Hirashima2024}. Contrary to the simple stellar population (SSP) approximation that is often used, which 
adopts the initial mass function (IMF) integrated yields \citep[e.g.,][]{Revaz2009, Okamoto2010, Hirai2015, Hirai2017a, Hirai2018, Hirai2019, Escala2018, Agertz2020}, star-by-star simulations can trace the ejecta of individual SNe. For example, \citet{Brauer2024} identified that the metallicity floor enriched by Population III SNe is, on average, [O/H] = $-4.0$, using their star-by-star simulation of high-$z$ galaxies.

The purpose of this Letter is to estimate the mono-enriched fraction of stars using our {dwarf galaxy} cosmological zoom-in simulation {resolving individual massive stars}. As an initial step to estimate this fraction, we focus on a dwarf galaxy with a simple chemical-enrichment history. The simulation adopts the model developed for Simulations Resolving Individual Stars (SIRIUS) project \citep{Hirai2021, Fujii2021b, Fujii2021a, Fujii2022a, Fujii2022b, Fujii2024}. 

This Letter is arranged as follows. Section \ref{sec:method} describes an updated stellar mass-assignment model. Section \ref{sec:results} shows the distributions and fractions of stars enriched by a single SN (mono-enriched) and those that are enriched by multiple SNe (multi-enriched). Section \ref{sec:discussion} compares our results with those from \citet{Hartwig2023}.

\section{Methods} \label{sec:method}
\subsection{Code}\label{sec:code}
We performed a cosmological zoom-in simulation of a dwarf galaxy with the $N$-body/smoothed particle hydrodynamics (SPH) code \textsc{asura}+\textsc{bridge} \citep{Fujii2021a}. Gravity was computed with a tree-based method \citep{Barnes1986}. Since we did not follow star cluster formation and evolution in this work, we turned off the BRIDGE scheme, which adopts the 6th-order Hermite integrator for short-range forces \citep{Fujii2007}. Hydrodynamics and models for galaxy formation were implemented with the \textsc{asura} code \citep{Saitoh2008, Saitoh2009, Hirai2022, Hirai2024a, Hirai2024b}. We adopted the density-independent SPH method \citep{Saitoh2013} to compute hydrodynamics and \textsc{cloudy} version 13.05 \citep{Ferland2013} for {atomic, metal-line, and molecular} cooling. We also implemented the ultra-violet background heating \citep{Haardt2012} and self-shielding models \citep{Rahmati2013}. {We did not include non-equilibrium chemistry for this study.}

Gas particles were allowed to form stars when the number density of hydrogen was higher than 100 cm$^{-3}$, the temperature was lower than 1000 K, and there was converging flow \citep{Hirai2021}. The code stochastically assigns the stellar mass to newly formed star particles (see Section \ref{sec:mass}). Once SNe occur, they distribute thermal energy and elements to the {128} nearest neighbor gas particles. Nucleosynthesis yields of core-collapse SNe \citep[CCSNe,][]{Nomoto2013}, type Ia SNe \citep[SNe Ia,][]{Seitenzahl2013}, and asymptotic giant branch (AGB) stars \citep{Cristallo2009, Cristallo2011, Cristallo2015} were compiled from the chemical-evolution library \citep[\textsc{celib},][]{Saitoh2017}. Ejected elements were diluted by the interstellar medium (ISM) following the diffusion equation. We adopt a scaling factor for the metal diffusion equation of 0.01, which is determined from the chemical abundances of MP stars in dwarf galaxies \citep{Hirai2017b}. {We have not included the effects of radiation from massive stars.}

\subsection{Stellar Mass Assignments}\label{sec:mass}
We updated the stellar mass-assignment model for newly formed star particles in \citet{Hirai2021} to sample the IMF in a simulation with $\sim$10 $M_{\sun}$ resolution. In this model, three kinds of star particles representing different ranges of the IMF were formed, adopting the Kroupa IMF \citep{Kroupa2001} from 0.1 to 120 $M_{\sun}$. {To prevent accumulating gas from unrealistically large and diffuse regions, we assumed gas mass should be accumulated in a region containing 5--10 times the star-particle mass ($m_{*}$) and within 3 pc, which was estimated from the threshold density of star formation \citep{Hirai2021}. These treatments suppress the formation of massive stars in our simulated dwarf galaxies. For this reason, we did not include stars more massive than 120 $M_{\sun}$, which were suggested by several simulations of Population III star formation \citep[e.g.,][]{Hirano2014, Hosokawa2016}.} Stars with $<6\,M_{\sun}$ were treated as SSP particles. We divided SSP particles into two mass ranges. The first type of SSP particles (SSP1) comprise stars with 0.1 $\,\leq\,m_*/M_{\sun}\,<1.3$, where $m_*$ is the stellar mass. We regarded SSP1 particles as purely chemically inert  
particles, i.e., these particles did not produce any stellar feedback. The second type of SSP particles (SSP2) comprises stars with 1.3 $\leq\,m_*/M_{\sun}\,<6.0$. We assumed these particles contribute to the feedback by AGB stars and SNe Ia. The IMF integrated yields of AGB stars and SNe Ia were produced from these particles. We treated stars with $m_*/M_{\sun}\,\geq6.0$ as star-by-star (SbS) particles, and assigned masses to them following \citet{Hirai2021}.
At the end of their lifetimes, these particles exploded as CCSNe and formed neutron 
stars {(13--40 $M_{\sun}$)}, or collapsed as white dwarfs {($<$13 $M_{\sun}$)} or black holes {($>$40 $M_{\sun}$), following \citet{Nomoto2013}}.

\subsection{Initial Conditions} \label{sec:ic}
A halo was selected from a cosmological simulation with a box size of (4 Mpc)$^3$ performed by \textsc{gadget-2} \citep{Springel2005}. We adopted the cosmological parameters of  $H_0=67.3$ km$\>$s$^{-1}$$\>$Mpc$^{-1}, \Omega_{\mathrm{m}}=0.32$, $\Omega_{\Lambda}=0.68$, and $\Omega_{\mathrm{b}}=0.049$ \citep{Planck2020}. A zoom-in initial condition was created by \textsc{music} \citep{Hahn2011}. The total number of particles in the zoom-in initial condition was 6.1$\,\times\,10^7$. We set the gravitational softening length as 20.2 pc for dark matter and 9.2 pc for gas particles. Dark matter and gas-particle masses in the zoom-in region were 102.3 $M_{\sun}$ and 18.9 $M_{\sun}$, respectively. {This resolution is sufficiently high to resolve individual massive stars and to estimate the mono-enriched fraction. However, it was still lower than the resolution where the galactic wind properties converged (see, e.g., \citealt{Hu2019}).} We computed this simulation over the redshift range $z$ = 100 to 6.5. At $z$ = 6.5, the dark matter halo mass, the stellar mass, and the median [Fe/H] were $1.5\times10^8$ $M_{\sun}$, $7.3\times10^5$ $M_{\sun}$, and $-2.2$, respectively. In total, 651,977 stars were formed. {We also assumed that the gas initially has zero metallicity.}

\section{Results} \label{sec:results}
\begin{figure}[ht!]
\epsscale{1.2}
\plotone{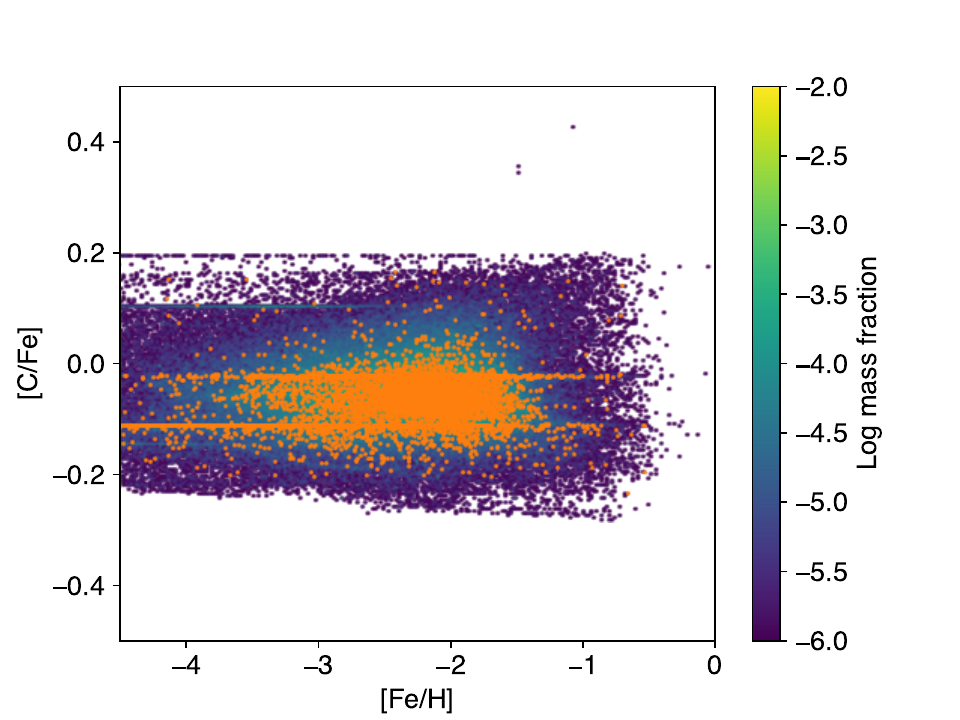}
\caption{[C/Fe], as a function of [Fe/H], at $z$ = 6.5. The color bar to the right encodes the mass fraction of stars contained in each metallicity bin, with a size of 0.02 $\times$ 0.004 dex$^2$. Orange dots show the mono-enriched stars.
\label{fig:cfe}}
\end{figure}

We identify mono-enriched stars using [C/Fe] ratios, motivated by \citet{Hartwig2023} and other work (e.g., \citealt{Vanni2023}, and references therein) that identified carbon as an indicator of mono-enriched stars. Since each CCSN has its own [C/Fe] ratio associated with its ejecta, mono-enriched stars should have the same [C/Fe] ratio as the yields of CCSNe, but the metallicity could differ depending on the degree of dilution. We regard a star as mono-enriched if the star has the same [C/Fe] ratio, within the numerical error ($\pm$10$^{-6}$ in this case), with the yields from previously occurring CCSNe. In this simulation, the contributions of AGB stars and SNe Ia are negligible. Figure \ref{fig:cfe} shows the simulated [C/Fe], as a function of [Fe/H], at $z$ = 6.5. As shown in this figure, mono-enriched stars have smaller dispersions than multi-enriched stars. Because not all CCSNe form stars directly from their ejecta, some multi-enriched stars have higher or lower [C/Fe] ratios than mono-enriched stars. {Some stars in constant [C/Fe] stripes are not labeled as mono-enriched. This is because our definition of mono-enriched is very tight (within $\pm10^{-6}$ from the SN yields). Although it appears constant, they are slightly affected by other SN ejecta.}

\begin{figure}[ht!]
\epsscale{1.2}
\plotone{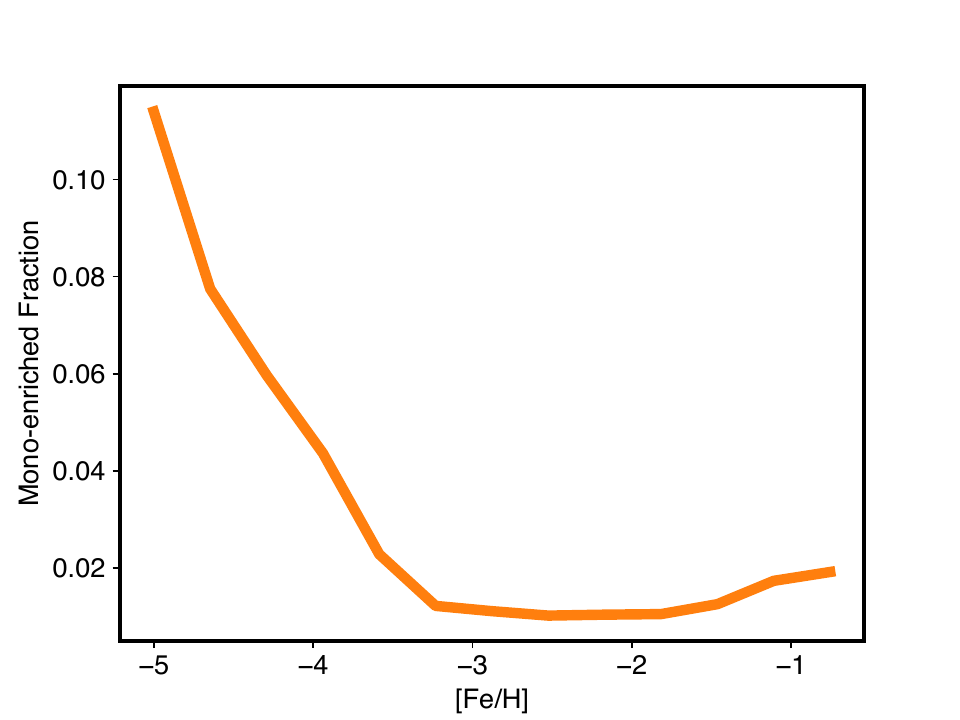}
\caption{The number fraction of mono-enriched stars, as a function of [Fe/H].
\label{fig:fraction}}
\end{figure}

Figure \ref{fig:fraction} highlights our main result. Using our {simulations}, we can compute the mono-enriched fraction as a function of [Fe/H]. As expected, the fraction exhibits an increasing trend toward lower metallicity for stars with [Fe/H] $< -2.5$. At [Fe/H] = $-5.0$, 11\% of the stars are mono-enriched. This result indicates that the ISM is less likely to be multiply enriched by CCSNe as metallicity decreases. For [Fe/H] $> -2.5$, the percentage of mono-enriched stars is 1\%. These relatively high-metallicity mono-enriched stars are directly formed from the ejecta of a CCSN without significant dilution into the ISM. Assuming a CCSN ejects 0.1 $M_{\sun}$ of Fe and the ejecta are mixed in a swept-up gas mass of $10^4\,M_{\sun}$ \citep{Cioffi1988}, the value of [Fe/H] is $-2.2$, indicating that stars with [Fe/H] $> -2.5$ can also be formed from a single SN ejecta. The slightly increased mono-enriched fraction for [Fe/H] $> -1.5$ is a numerical artifact due to the small number of stars in this metallicity range. {This result is not altered if we only plot low-mass stars that can survive to $z$ = 0.}

\begin{figure}[ht!]
\epsscale{1.2}
\plotone{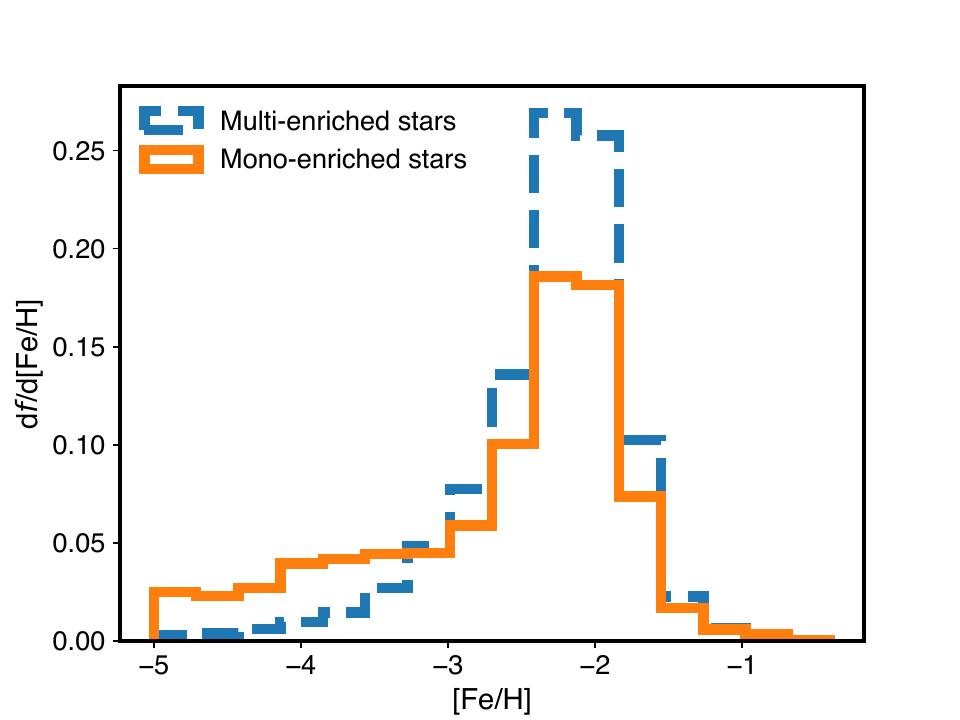}
\caption{The metallicity distribution functions of multi-enriched (blue-dashed line) and mono-enriched (orange-solid line) stars at $z$ = 6.5.
\label{fig:mdf}}
\end{figure}

Figure \ref{fig:mdf} shows metallicity distribution functions (MDFs) of mono- and multi-enriched stars. From inspection, the mono-enriched MDF exhibits an enhanced low-metallicity tail, while the multi-enriched MDF smoothly decreases as [Fe/H] decreases. This feature reflects the increasing trend of the mono-enriched fraction toward lower metallicity for [Fe/H] $< -2.5$ (Figure \ref{fig:fraction}). For higher metallicity, the MDFs are indistinguishable. The median [Fe/H] for mono- and multi-enriched stars are $-$2.5 and $-$2.2, respectively.

\begin{figure}[ht!]
\epsscale{1.2}
\plotone{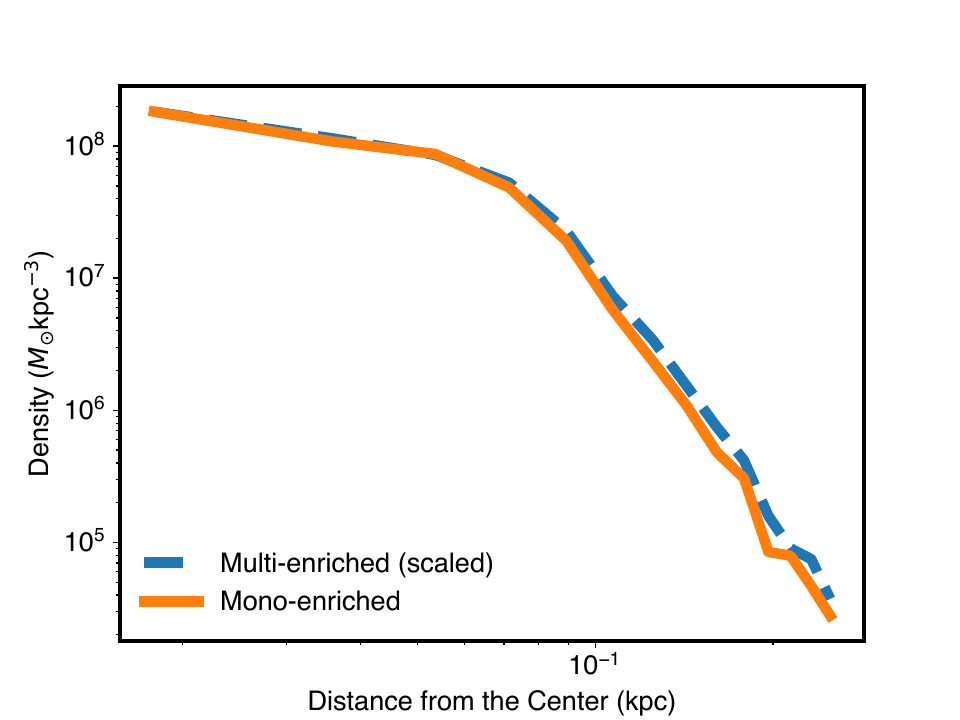}
\caption{The {density profiles} of multi-enriched (blue{-dashed line}) and mono-enriched (orange{-solid line}) stars at $z$ = 6.5. {The density profile of multi-enriched stars is multiplied by 0.017 to be scaled to that of mono-enriched stars.}
\label{fig:spatial}}
\end{figure}

\begin{figure}[ht!]
\epsscale{1.2}
\plotone{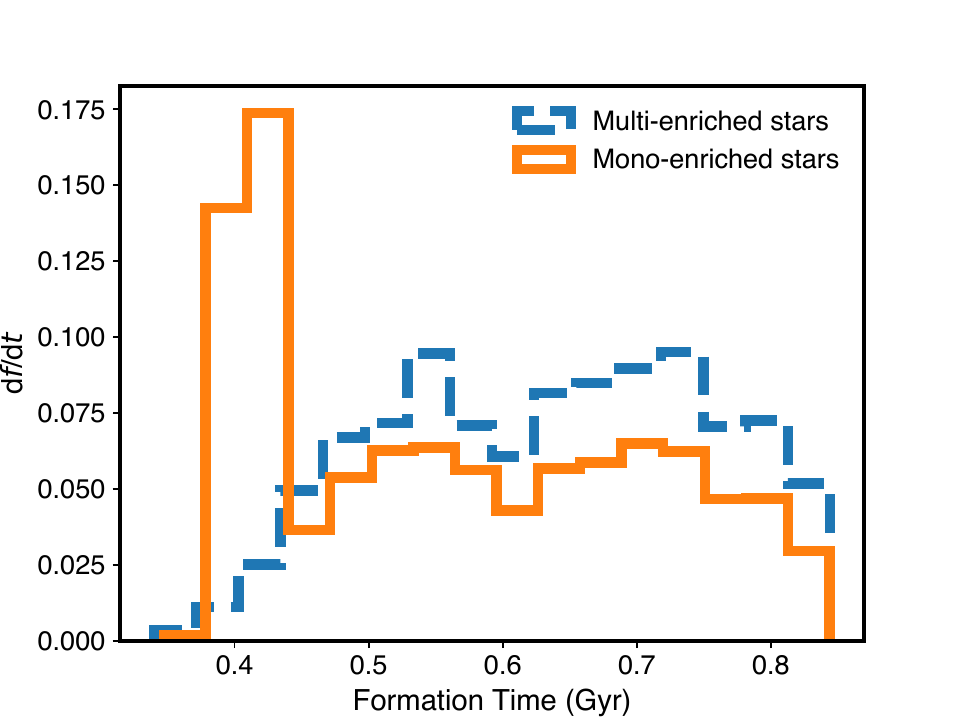}
\caption{The formation times of multi-enriched (blue-dashed line) and mono-enriched (orange-solid line) stars.
\label{fig:time}}
\end{figure}

The spatial distribution and formation times of mono-enriched stars reflect the inside-out formation of the simulated dwarf galaxy. Figure \ref{fig:spatial} depicts the {density profiles} of mono-enriched stars. As shown in this figure, {the profile is almost identical, but multi-enriched stars have slightly higher density in the outskirts than mono-enriched stars.} This is because the largest fraction of mono-enriched stars are formed in the earliest phases of galaxy formation (Figure \ref{fig:time}). In the early phase, stars are mainly formed in the galactic center, where the gas density is the highest. In later phases, the star-formation region extends to the surrounding region, but the ISM tends to be mixed with multiple SNe. From inspection of Figure \ref{fig:time}, mono-enriched stars are continuously formed after the initial burst at 0.4 Gyr. These stars are also formed near the galactic center, where high-density gas allows spontaneous star formation following a SN explosion.

\section{Discussion and Summary} \label{sec:discussion}
For the first time, this study has computed the mono-enriched fraction, as a function of [Fe/H], directly from a simulation. We have shown an increasing trend in this fraction toward lower metallicity for [Fe/H] $< -2.5$. At [Fe/H] = $-$5, 11\% of the stars are mono-enriched. In this simulated dwarf galaxy, 1.5\% of the stars are mono-enriched, suggesting that MP stars are mainly formed from gas enriched by multiple SNe.

The increasing mono-enriched fraction toward lower metallicity is also seen in the observation-based machine learning estimate \citep{Hartwig2023}. They pointed out that not all stars were mono-enriched at [Fe/H] $\sim -4.5$. Since our approach is completely different from their method, Figure \ref{fig:fraction} is an independent confirmation of the increasing trend of the mono-enriched fraction toward lower metallicity.

The major difference between these studies is the predicted fraction of mono-enriched stars. For EMP stars, the mono-enriched fraction in this study is 5.0\%, while for those in \citet{Hartwig2023} it is 31.8\%. Since this study adopts the strict condition that stellar [C/Fe] ratios must be the same as the yields of CCSNe within $\pm 10^{-6}$, our estimate is expected to be lower than observation-based estimates. Typical observational uncertainties for [C/Fe] ratios are $\pm$0.2 dex. Observationally, the chemical abundances of many of the multi-enriched stars found in this study are indistinguishable from those of mono-enriched stars. Because of the observational uncertainties that would increase the mono-enriched fraction, the difference between our study and \citet{Hartwig2023} appears reasonable.

Star-formation histories {and the change of the IMF as a function of the metallicity \citep[e.g.,][]{Chon2021} could also affect the estimates of the} mono-enriched fraction. Due to the limitations of computational resources, we only computed a single cosmological zoom-in simulation. Future simulations with different initial conditions will show how star-formation histories {and the IMF} affect a galaxy's mono-enriched fractions. The adopted metal-diffusion coefficient 
could also affect our estimate. However, this parameter does not affect the predicted increasing trend toward lower metallicity of the mono-enriched fraction, because the metallicity does not affect the value of the diffusion coefficient.

From the discussion above, our simulation suggests that the mono-enriched fraction is larger at lower metallicity for stars with [Fe/H] $< -2.5$. This result is the first quantitative estimate of the mono-enriched fraction from a galaxy simulation. Our work indicates that we could clearly benefit from increasing the number of EMP stars with measured chemical abundances. \citet{Huang2024} reported photometric chemical abundances of five million stars{, including 25,704 EMP stars}. Assuming the average mono-enriched fraction of {5\% for EMP stars} in this work, the data contain at least {1000} mono-enriched stars. Spectroscopic follow-up of such a large sample is crucial to identify mono-enriched stars. 

{This Letter focuses on computing the mono-enriched fraction. The next step is to construct models that can reproduce observed chemical abundances of dwarf galaxies. For example, finding a condition to form CEMP stars, which are not formed in this simulation, is of interest. \citet{Brauer2024} have shown that Population III supernovae could form CEMP stars with their cosmological simulation adopting a top-heavy Population III IMF. Because we did not adopt the top-heavy IMF, our simulations would suppress the formation of CEMP stars. CEMP stars could be also formed from faint SNe \citep[e.g.,][]{Ishigaki2014}, rapidly rotating massive stars \citep[e.g.,][]{Meynet2006}, or the mass transfer between binaries \citep[e.g.,][]{Suda2004}.}

Recently, there have been a number of massive observational projects for the discovery of MP stars, as described in the Introduction.  The soon-to-begin Subaru Prime Focus Spectrograph (PFS) \citep{Takada2014} will also be able to identify large numbers of potentially mono-enriched stars, as a portion of this effort is directed at spectroscopic observations of individual dwarf galaxy satellites of the Milky Way. These surveys should contain at least 5\% of EMP stars that are mono-enriched. From inspection of Figure \ref{fig:spatial}, we can efficiently find mono-enriched stars by targeting EMP stars near the center of dwarf galaxies. {Among these stars, signatures of Population III supernovae, such as pair-instability supernovae \citep[e.g.,][]{Takahashi2018} and rapidly rotating massive stars \citep[e.g.,][]{Tsiatsiou2024}, could be detected.} Comparison with star-by-star simulations and these survey data will greatly improve our understanding of SNe nucleosynthesis occuring in the earliest phase of galaxy formation.

\begin{acknowledgments}
{We are grateful to an anonymous referee for suggestions, which improved the clarity of our presentation.} This work was supported in part by JSPS KAKENHI Grant Numbers JP22KJ0157, JP21H04499, JP21K03614, JP22H01259, JP21H05448, MEXT as ``Program for Promoting Researches on the Supercomputer Fugaku" (Structure and Evolution of the Universe Unraveled by Fusion of Simulation and AI; Grant Number JPMXP1020230406), JICFuS, grants PHY 14-30152; Physics Frontier Center/JINA Center for the Evolution of the Elements (JINA-CEE), and OISE-1927130: The International Research Network for Nuclear Astrophysics (IReNA), awarded by the US National Science Foundation. Numerical computations and analysis were carried out on Cray XC50 and computers at the Center for Computational Astrophysics, National Astronomical Observatory of Japan and the  Yukawa Institute Computer Facility. This research has made use of NASA's Astrophysics Data System.
\end{acknowledgments}
\vspace{5mm}
\software{
ASURA+BRIDGE \citep{Fujii2021a},
CELib \citep{Saitoh2017},
          Cloudy \citep{Ferland2013},
          Gadget-2 \citep{Springel2005},
          MUSIC \citep{Hahn2011}}
\vfill\eject
\bibliography{ms}{}
\bibliographystyle{aasjournal}
\end{document}